\documentclass[twocolumn,showpacs,preprintnumbers,amsmath,amssymb]{revtex4}
\usepackage{graphicx}
\usepackage{tabularx}
\usepackage{dcolumn}
\usepackage{bm}
\usepackage{mathrsfs}
\usepackage{mathtools}
\usepackage{amsmath}
\usepackage{booktabs}

\begin{document}
\preprint{APS/123-QED}

\title{Structure of the neutron mid-shell nuclei $^{111,113}_{\quad \ 47}$Ag$_{64,66}$}

\author{S.~Lalkovski$^1$\footnote{e-mail: stl@phys.uni-sofia.bg}, 
E.~A.~Stefanova$^2$, S.~Kisyov$^3$, A.~Korichi$^4$, D.~Bazzacco$^5$, 
M.~Bergstr\"om$^6$, A.~G\"orgen$^7$, B.~Herskind$^6$, H.~H\"ubel$^8$, A.~Jansen$^8$, 
T.L.~Khoo$^9$, T.~Kutsarova$^2$, A.~Lopez-Martens$^4$, A.~Minkova$^1$, 
Zs.~Podoly\'ak$^{10}$, G.~Sch\"onwasser$^8$, O.~Yordanov$^2$}

\affiliation{
$^1$Department of Nuclear Engineering, Faculty of Physics, University of Sofia ''St. Kl. Ohridski'', Sofia 1164, Bulgaria\\
$^2$Institute for Nuclear Research and Nuclear Energy, Bulgarian Academy of Science, Sofia 1784, Bulgaria\\
$^3$''Horia Hulubei'' National Institute for Physics and Nuclear Engineering, RO-077125 Bucharest, Romania \\
$^4$CSNSM Orsay, IN2P3/CNRS, F-91405, France\\
$^5$INFN, Sezione di Padova, I-35131 Padova, Italy\\
$^6$The Niels Bohr Institut, Blegdamsvej 17, DK-2100 Copenhagen, Denmark\\
$^7$Department of Physics, Faculty of Mathematics and Natural Sciences, University of Oslo, Norway\\
$^8$Helmholtz-Institut f\"ur Strahlen -und Kernphysik, Universit\"at Bonn, Nussallee 14-16, D-53115 Bonn, Germany\\
$^9$Physics Division, Argonne National Laboratory, Argonne, Illinois 60439, USA\\
$^{10}$ Department of Physics, University of Surrey, Guildford GU27XH, UK\\
}

\date{\today}

\begin{abstract}
$^{111}$Ag and $^{113}$Ag were produced in induced fission reaction, where yrast 
and near-yrast states were populated. To interpret the new data the Interacting 
Boson-Fermion model was used. A good agreement with the experimental data is 
achieved, suggesting that the two Ag nuclei have a well developed collectivity, 
superimposed on $\pi g_{9/2}^{-3}$ excitations previously observed throughout 
the entire isotopic chain.
\end{abstract}

\maketitle
\section{Introduction}
Silver nuclei present an excellent ground for testing of different theoretical 
models. Being three proton holes away from the Sn nuclei, they represent
a good test case for the Nuclear Shell model \cite{GM48}. Indeed, the $j-1$ 
anomaly, observed in the low-energy part of Ag spectra, is often interpreted as 
arising from three-hole clusters \cite{Ki66} -- a direct derivative 
from the Shell model \cite{He94}. Within that approach, however, the $M1$ 
transitions between members of the same multiplet are forbidden \cite{PVI} which, 
to certain extent, coincides with the experimental observation \cite{Sv76}. 
Detailed shell model calculations \cite{La13}, however, seem to fail in 
describing those states. Neither their ordering, nor the $(j,j-1)$ energy gap is 
well reproduced, which was assumed to arise from enhanced $p-n$ interaction. 

Attempts to explain the structure of the low-lying states in Ag nuclei were also 
made in the framework of different collective and algebraic models \cite{HP86}. 
Indeed, already in the early 60's de-Shalit \cite{Sh61} demonstrated that the 
low-lying negative-parity states in Ag isotopes represent core excitations, 
weakly coupled to the odd unpaired particle. These studies were followed by the 
Cluster-Vibration model developed \cite{Pa73} and applied for an 
extensive set of levels in the Ag nuclei. Quasi-particle-plus rotor model 
calculations were performed  in Refs.~\cite{Po79, Ze86} and Interacting 
Boson-Fermion Model calculations in Ref.~\cite{Ka81, Ro90}. It was pointed out 
\cite{HP86} that, in addition to magnetic moments and electromagnetic transition 
strengths, a reasonable explanation of the $j-1$ anomaly can be achieved by using 
enhanced quadrupole residual interaction.

Although the neutron mid-shell Ag nuclei are of paramount importance for 
testing of various models, the experimental data is often scarce. In particular,
little is known about the yrast states in the mid-shell $^{111}$Ag \cite{Bl09} 
and $^{113}$Ag \cite{Bl10}. To fill in the gap, we report on new data from 
induced fission reaction. The extended level schemes are analyzed by using 
IBFM-1 calculations.

\section{Experimental details}

\begin{figure*}[ht]
\rotatebox{90}{\scalebox{0.5}[0.5]{\includegraphics{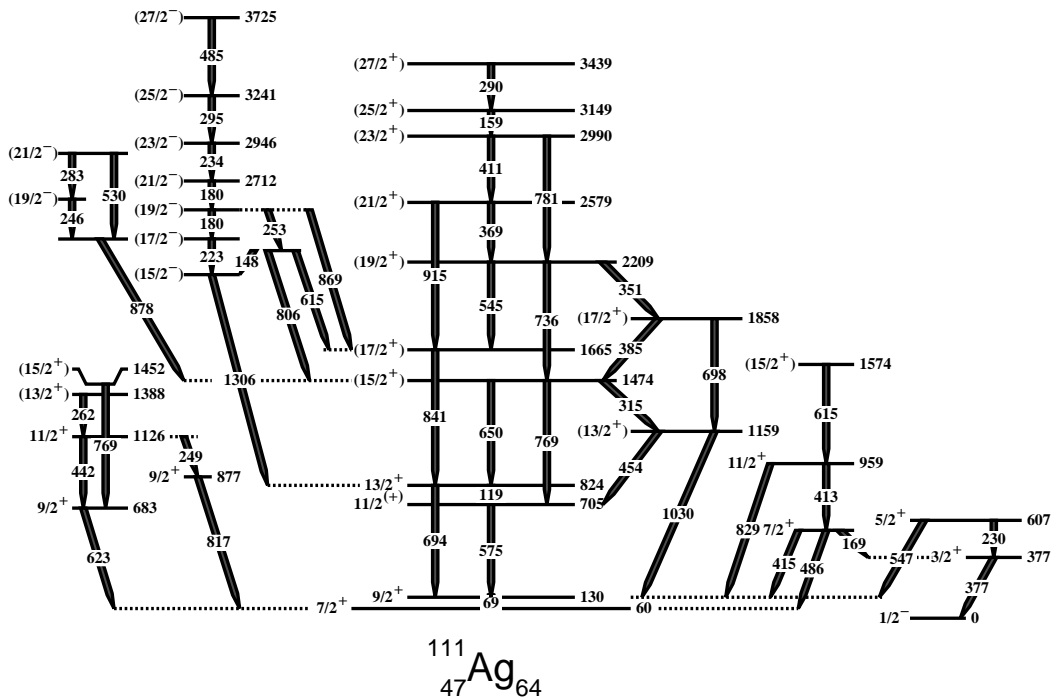}}}
\caption[]{\label{111Aglev}$^{111}$Ag partial level scheme.}
\end{figure*}

$^{111,113}$Ag were produced in induced fission reaction. The $^{30}_{14}$Si 
beam was accelerated up to $E=142$ MeV by the XTU tandem accelerator at the 
Legnaro National Laboratory, and impinged on a 1.15 mg/cm$^2$ thick 
$^{168}_{\ 68}$Er target. To stop the recoil fission fragments, 
the target was deposited on a 9 mg/cm$^2$ gold backing. Gamma-rays, emitted from 
the excited nuclei, were detected by the  EUROBALL III multi-detector array 
\cite{Si97}, comprising 30 single HPGe detectors, 26 Clover and 15 Cluster 
detectors with anti-Compton shields. Triple $\gamma$ ray coincidences were 
recorded. The data was sorted in $E_\gamma-E_\gamma-E_\gamma$ cubes and analyzed 
by using the RADWARE software \cite{Ra95}.
\begin{figure}[ht]
\rotatebox{0}{\scalebox{0.3}[0.3]{\includegraphics{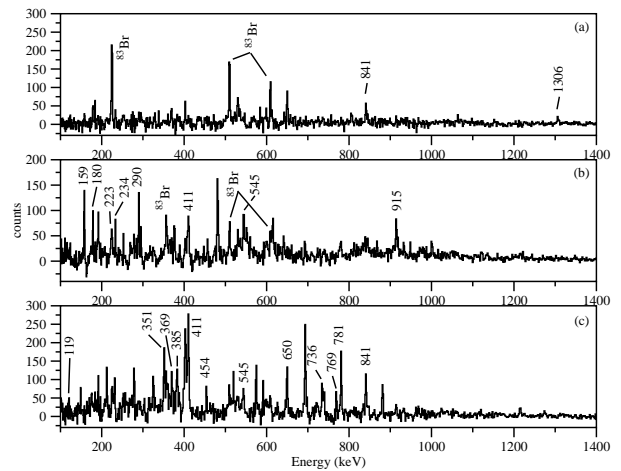}}}
\caption[]{\label{111Agspec}Sample coincidence spectra for $^{111}$Ag, gated on
356-keV $\gamma$ ray in $^{83}$Br and 694-keV $\gamma$ ray in $^{111}$Ag (a), and
on 694- and 841-keV $\gamma$ rays (b), and 290- and 159-keV $\gamma$ rays (c), 
respectively.}
\end{figure}

\begin{table}[ht]
\caption{\label{111Pd_tabl}Data set of $^{111}$Ag levels and $\gamma$-rays as
observed in the present work, but also known from the literature \cite{Bl09}. Level 
energies, $E_i$ and $E_f$ in keV, are obtained after a least-squares fit to the 
energies of the $\gamma$ rays $E_\gamma$ connecting the respective initial and 
final states. $BR_\gamma$ are the gamma-decay branching ratios. With few exceptions 
discussed in the text, the spin/parity assignments $J^\pi$ are from Ref.~\cite{Bl09}.}
\begin{tabular}{llllllllll}
\hline\hline
$E_i$ & $J_i^\pi$    &   $E_f$ & $J_f^\pi$    & $E_\gamma$ & $BR_\gamma$ \\
\hline
0       & $1/2^-$    	 &         &              &            &        \\
60.1    & $7/2^+$    	 &         &              &            &        \\
129.6   & $9/2^+$     	 & 60.1    & $7/2^+$      &  68.8      &        \\
376.6   & $3/2^+$    	 & 0       & $1/2^-$      &  376.6     &        \\
545.4   & $7/2^+$    	 & 376.6   & $3/2^+$      &  168.9     &  23 (2)\\
        &            	 & 129.6   & $9/2^+$      &  415.4     &  100   \\
        &            	 & 60.1    & $7/2^+$      &  486.0     &  16 (2)\\
607.0   & $5/2^+$    	 & 376.6   & $3/2^+$      &  230.3     &        \\
        &            	 & 60.1    & $7/2^+$      &  547.0     &        \\
683.3   & $9/2^+$    	 & 60.1    & $7/2^+$      &  623.3     &        \\
704.9   & $11/2^{(+)}$   & 129.6   & $9/2^+$      &  575.2     &        \\ 
823.9   & $13/2^+$   	 & 704.9   & $11/2^{(+)}$ &  118.9     &  $<5$  \\
        &            	 & 129.6   & $9/2^+$      &  694.3     & 100    \\
876.9   & $9/2^+$    	 & 60.1    & $7/2^+$      &  816.8     &        \\
958.5   & $11/2^+$   	 & 544.4   & $7/2^+$      &  413.3     &        \\
        &            	 & 129.6   & $9/2^+$      &  828.7     &        \\
1125.7  & $11/2^+$    	 & 876.9   & $9/2^+$      &  248.7     &        \\
        &            	 & 683.3   & $9/2^+$      &  442.4     &        \\
1159.4  & $(13/2^+)$ 	 & 704.9   & $11/2^{(+)}$ &  454.2     &        \\  
        &            	 & 129.6   & $9/2^+$      &  1030.0    &        \\
1388.3  & $(13/2^+)$ 	 & 1125.7  & $11/2^+$     &  262.6     &        \\
1452.1  & $(15/2^+)$	 & 683.3   & $9/2^+$      &  768.8     &        \\
1474.0  & $(15/2^+)$ 	 & 1159.4  & $(13/2^+)$   &  314.8     & 12 (2) \\
        &            	 & 823.9   & $13/2^+$     &  650.3     & 100    \\
        &            	 & 704.9   & $11/2^{(+)}$ &  769.4     & 55 (4) \\
1573.6  & $(15/2^+)$     & 958.5   & $11/2^+$     &  615.1     &        \\
1664.5  & $(17/2^+)$     & 823.9   & $13/2^+$     &  840.9     &        \\
2130.6  & $(15/2^-)$     & 823.9   & $13/2^+$     & 1306.1     &        \\
2352.8  & $(17/2^-)$     & 2130.6  & $(15/2^-)$   & 222.6      &        \\
        &                & 1474.0  & $(15/2^+)$   & 878.4      &        \\
\hline\hline
\end{tabular}\\
\end{table}
\begin{table}[ht]
\caption{\label{111Pd_tablc} Data set of $^{111}$Ag levels and gamma rays 
observed for the first time in the present study. Continues from Table~\ref{111Pd_tabl}.}
\begin{tabular}{llllllllll}
\hline\hline
$E_i$ & $J_i^\pi$    &   $E_f$ & $J_f^\pi$    & $E_\gamma$ & $BR_\gamma$ \\
\hline

1858.2  & $(17/2^+)$   & 1474.0  & $(15/2^+)$     &  384.6     & 100    \\
        &              & 1159.4  & $(13/2^+)$     &  698.3     & 55 (7) \\
2209.4  & $(19/2^+)$   & 1858.2  & $(17/2^+)$   & 351.1   & 60 (5) \\ 
        &              & 1664.5  & $(17/2^+)$   & 544.5   & 78 (6) \\
        &              & 1474.0  & $(15/2^+)$   & 735.6   & 100    \\
2279.9  & $(17/2^+)$   & 2130.6  & $(15/2^-)$   & 148.1   & 70 (3) \\
        &              & 1664.5  & $(17/2^+)$   & 615.0   & 100    \\
        &              & 1474.0  & $(15/2^+)$   & 806.4   &  38 (3) \\
2532.7  & $(19/2^-)$   & 2352.8  & $(17/2^-)$   & 179.8   &        \\
        &              & 2279.9  & $(17/2^+)$   & 252.7   &        \\
        &              & 1664.5  & $(17/2^+)$   & 868.7   &        \\
2579.2  & $(21/2^+)$   & 2209.4  & $(19/2^+)$   & 369.2   & 13 (2) \\
        &              & 1664.5  & $(17/2^+)$   & 914.9   & 100    \\
2599.1  & $(19/2^+)$   & 2352.8  & $(17/2^-)$   & 246.1   &        \\
2712.3  & $(21/2^-)$   & 2532.7  & $(19/2^-)$   & 179.6   &        \\
2882.6  & $(21/2^+)$   & 2599.1  & $(19/2^+)$   & 283.3   &        \\
        &              & 2352.8  & $(17/2^-)$   & 530     &        \\ 
2946.0  & $(23/2^-)$   & 2712.3  & $(21/2^-)$   & 233.7   &        \\
2990.1  & $(23/2^+)$   & 2579.2  & $(21/2^+)$   & 410.5   & 100    \\
        &              & 2209.4  & $(19/2^+)$   & 781.1   & 79 (4) \\
3148.9  & $(25/2^+)$   & 2990.1  & $(23/2^+)$   & 158.8   &        \\
3240.9  & $(25/2^-)$   & 2946.0  & $(23/2^-)$   & 294.9   &        \\
3438.7  & $(27/2^+)$   & 3148.9  & $(25/2^+)$   & 289.8   &        \\  
3725    & $(27/2^-)$   & 3241    & $(25/2^-)$   & 484.5   &        \\
\hline\hline
\end{tabular}\\
\end{table}

The compound nucleus, produced in this reaction, is $^{198}_{\ 72}$Pt. The 
dominant reaction channel is the fusion/evaporation reaction leading to 
$^{194,195}_{\quad \ \ \ 82}$Pb \cite{Ku07}. The induced fission reaction 
represents only a small fraction of the total cross section. In such reactions, 
the proton evaporation is highly suppressed and the sum of the fragments' atomic 
numbers is equal to the atomic number of the fissioning system. Hence, the 
most populated complementary fragments of $^{111}$Ag and $^{113}$Ag are 
$^{83}_{35}$Br and $^{81}_{35}$Br, respectively. More details about this 
experiment, and other fission products produced in it, are published in 
Refs.~\cite{La07,St12,La14}. 

\subsection{$^{111}$Ag}
The partial level scheme, obtained from cross coincidences between $^{111}$Ag 
and its complementary fragment $^{83}$Br, is presented in Fig.~\ref{111Aglev}. 
Once coincident transitions are established to belong to $^{111}$Ag, they were 
used to further extend the level scheme. The procedure used in the present work 
is similar to the one used to build the positive-parity bands placed on top of 
the ground states of $^{107,109}$Pd \cite{St12} and $^{105}$Ru \cite{La14}. 
Sample coincidence spectra for $^{111}$Ag are shown in Fig.~\ref{111Agspec}. 
Gamma-ray energies $E_\gamma$ and branching ratios $BR_\gamma$ are listed in 
Table~\ref{111Pd_tabl} and \ref{111Pd_tablc} along with energies of the initial 
$E_i$ and final $E_f$ states, obtained from a least-squares fit to $E_\gamma$. 
Table~\ref{111Pd_tabl} presents the levels known prior to our study. Indeed, 
$^{111}$Ag \cite{Bl09} was previously studied from $^{111}$Pd $\beta ^-$ decay 
\cite{BK88}, $^{109}$Ag(t,p) \cite{An77}, $^{110}$Pd($^3$He,d) \cite{AK77}, 
$^{110}$Pd($^3$He,pn$\gamma$) \cite{Ze87}, and $^{112}$Cd(d,$^{3}$He) \cite{We76} 
reactions. In addition to the levels presented in Table~\ref{111Pd_tabl} many
other non-yrast positive and negative parity were previously observed \cite{Bl09}.
Table~\ref{111Pd_tablc} lists the levels and gamma-rays observed for the first 
time in the present work.

Due to the poor statistics in the present study no angular correlation nor 
angular distribution measurements were performed. Therefore, the spin and 
parities in the present work are based on: the spin and parities already assigned 
in Refs.~\cite{Bl09,Bl10}; the analogy with $^{107}$Ag, where angular correlation 
measurements were performed; and on the systematics \cite{La13}. The new
transitions are assumed to be of dipole and quadrupole multipolarity only.

In Ref.~\cite{Bl09}, based on reaction data and angular distribution measurements, 
spin and parity assignments have been made to all states below the 1574-keV state. 
In contrast to \cite{Bl09} where (11/2) is assigned to the 1388-keV level we 
tentatively assign $(13/2^+)$ assuming that in the induced fission experiments 
spins increase with the energy. Based on the same argument the spin/parity 
assignments were made to the 1159-, 1452-keV levels. The spin/parity assignments
to the 1474-, 1665-keV levels are also based on the afore mentioned argument and 
the systematics. The structure built of low energy transitions on top of the 
2131-keV state, is similar to the structure observed on top of the 2298-keV state 
in $^{107}$Ag. In $^{111}$Ag, as in $^{107}$Ag, the sequence decays to the yrast 
$13/2^+$ state via 1.3-MeV  transition. Therefore, we tentatively assume this 
sequence to be a $\Delta I=1$ band of negative-parity states. 

The sequence, built on top of the $7/2^+$ state at 60 keV, is extended up to 
3439 keV. This band-like structure resembles the sequence observed in $^{107}$Ag 
\cite{Es97} and is consistent with the systematic trend of the low-lying 
positive-parity yrast states in the Ag isotopic chain \cite{La13}.

\begin{figure}[ht]
\rotatebox{-90}{\scalebox{0.35}[0.35]{\includegraphics{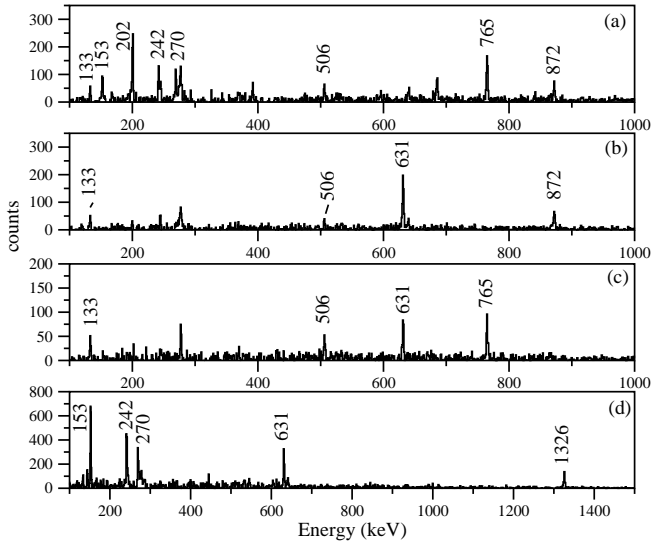}}}
\caption[]{\label{113Agspec}Sample coincidence spectra for $^{113}$Ag, representing 
the 95$\gamma$-631$\gamma$ (a), 95$\gamma$-765$\gamma$ (b), 
95$\gamma$-872$\gamma$ (c), and 202$\gamma$-201$\gamma$ coincidences (d).}
\end{figure}

\begin{figure}[ht]
\rotatebox{0}{\scalebox{0.35}[0.35]{\includegraphics{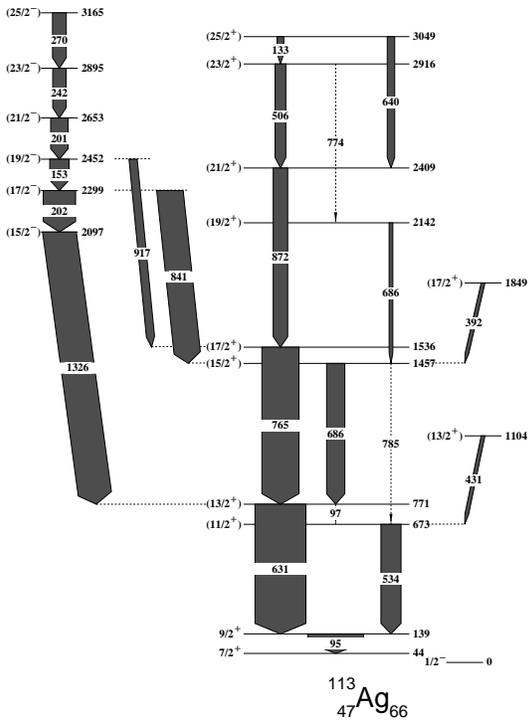}}}
\caption[]{\label{113Aglev}$^{113}$Ag partial level scheme.}
\end{figure}

\begin{table}[ht]
\caption{\label{113Pd_tabl}Data set of $^{113}$Ag levels and gamma rays as 
observed in the present work. Level energies $E_i$ and $E_f$ are obtained after
a least-squares fit to the $\gamma$-ray energies $E_\gamma$ in keV. Relative 
$\gamma$ ray intensities ($I_\gamma$) are normalized to $I_{765}=100\%$. Except 
for the levels, denoted with an $^a$ symbols which were known prior to this 
study \cite{Bl10}, the spin/parity assignments $J^\pi$ to the rest of the levels 
are from this work.}
\begin{tabular}{llllllllll}
\hline\hline
$E_i$ & $J_i^\pi$&  $E_f$& $J_f^\pi$ & $E_\gamma$ & $I_\gamma$\\
\hline
0$^a$       & $1/2^-$    &       &             &       &    \\
44$^a$      & $7/2^+$    &       &             &       &    \\
139.1$^a$   & $9/2^+$    & 44    &  $7/2^+$    & 95.1  & 94 \\
673.1$^a$   & $(11/2^+)$ & 139.1 &   $9/2^+$   & 534.2 & 54 \\
770.7       & $(13/2^+)$ & 139.1 &   $9/2^+$   & 631.4 & 137\\
            &            & 673.1 &  $(11/2^+)$ & 97.1  & 39 \\
1104.1      & $(13/2^+)$ & 673.1 & $(11/2^+)$  & 431   &    \\
1457.0      & $(15/2^+)$ & 770.7 & $(13/2^+)$  & 685.6 & 49 \\
            &            & 673.1 & $(11/2^+)$  & 784.6 & 20 \\
1536.0      & $(17/2^+)$ & 770.7 & $(13/2^+)$  & 765.3 & 100\\
1849.0      & $(17/2^+)$ & 1457.0& $(15/2^+)$  & 392   &    \\
2096.8      & $(15/2^-)$ & 770.7 & $(13/2^+)$  & 1326.0& 104\\
2142.3      & $(19/2^+)$ & 1457.0& $(15/2^+)$  & 686   &    \\
2298.8      & $(17/2^-)$ & 2096.8& $(15/2^-)$  & 202   & 87 \\
            &            & 1457.0& $(15/2^+)$  & 841.2 & 71 \\
2408.7      & $(21/2^+)$ & 1536.0& $(17/2^+)$  & 872.1 & 39 \\
2452.1      & $(19/2^-)$ & 2298.8& $(17/2^-)$  & 152.6 & 52 \\
            &            & 1457.0& $(15/2^+)$  & 916.8 & 22 \\
2652.7      & $(21/2^-)$ & 2452.1& $(19/2^-)$  & 200.6 & 47 \\
2894.9      & $(23/2^-)$ & 2652.7& $(21/2^-)$  & 242.2 & 29 \\
2915.7      & $(23/2^+)$ & 2408.7& $(21/2^+)$  & 506.2 & 29 \\
3049.0      & $(25/2^+)$ & 2915.7& $(23/2^+)$  & 133.2 & 18 \\
            &            & 2408.7& $(21/2^+)$  & 640.4 & 20 \\
3164.5      & $(25/2^-)$ & 2894.9& $(23/2^-)$  & 269.6 & 36 \\
\hline\hline
\end{tabular}\\
\end{table}

\subsection{$^{113}$Ag}
$^{113}$Ag \cite{Bl10} was studied previously via $\beta ^-$ decay of $^{113}$Pd.
Many low-spin states of positive and negative parities were observed, but the 
level scheme is incomplete given that $Q_{\beta -}=3.4$ MeV and only levels up 
to 0.783 MeV were experimentally observed.

Being further away from the line of $\beta ^-$-stability than $^{111}$Ag, $^{113}$Ag 
is difficult to populate via light particle transfer reactions and fusion/evaporation 
reactions. Also, it is not well produced in spontaneous fission. Thus, the 
experiment described in the present work opens a new opportunity to study this
particular nucleus, given that the fissioning $^{198}$Pb system is lighter then 
any of the spontaneous fissioning sources traditionally used to populate 
nuclei on the neutron-rich side of the $\beta^-$ stability line. 

In all complementary Br fragments, lines with energies of 95, 631 and 534 keV 
were observed. The 95- and 534-keV lines were known to belong to $^{113}$Ag 
\cite{Bl10}. They were used in the present work to identify the most populated 
complementary fragment of $^{113}$Ag, which is $^{81}$Br. Coincidences between 
these 631- and 534-keV transitions and the most intense transitions in $^{81}$Br 
were analysed to deduce the other yrast states in $^{113}$Ag. The higher lying 
states in $^{113}$Ag were established from coincidences with the most intense 
transitions already assigned to $^{113}$Ag. Sample coincidence spectra are 
displayed in Fig.~\ref{113Agspec} and the new level scheme is shown in 
Fig.~\ref{113Aglev}. 

It has to be noted, that the 431-keV transition is also in 
coincidence with the transitions from the sequence on top of the 2097-keV level. 
However, no link between the two structures was observed, suggesting that the 
2097-keV level decays via multiple week transitions to several intermediate states.

The spin and parity assignments to the levels are based on the values adopted in 
Ref.~\cite{Bl10} for the lowest-lying states, on systematics \cite{La13}, and on 
analogy with the $^{115}$Ag level scheme. Again, only dipole and quadrupole type
of transitions are assumed.

\section{Discussion}

\subsection{IBM-1 calculations}
To interpret the new data, Interacting Boson-Fermion Model (IBFM) \cite{Ia79,Ia91}
calculations were performed for $^{111,113}$Ag. The model describes the excited
states of odd-A nuclei via a coupling of the last unpaired fermion to the 
even-even bosonic core. In the present work, the cadmium nuclei $^{112,114}$Cd
are considered to be the even-even cores of $^{111,113}$Ag. Core excited 
states were calculated by using the IBM-1 model \cite{Ia74,Ar75,Ca88,Pf98} 
within its extended consistent-$Q$ formalism (ECQF) \cite{Li85,Bu86}, 
where the model Hamiltonian can be written as 
\begin{equation}
H=\varepsilon n_d-\kappa Q^2 -\kappa' L^2 .
\end{equation}
Here, $$n_d=\sqrt{5}T_0$$ is the number of $d$-bosons operator. The total number 
of bosons $N = n_s + n_d$ is then taken as half the number of valence particles 
or holes, counted from the nearest closed-shell gap \cite{Ca90}. This version of 
the IBM does not distinguish protons from neutrons. 

In IBM-1, the angular momentum operator is defined as
$$L=\sqrt{10}T_1$$ 
and the quadrupole operator as $$Q=(d^{\dagger} s+s^{\dagger}\widetilde{d})
+\chi (d^{\dagger}\widetilde{d})^{(2)}=(d^{\dagger} s+s^{\dagger}\widetilde{d})
+\chi T_2 \ ,$$ where
$$\widetilde{d_\mu}=(-1)^{\mu}d_{-\mu} $$

\begin{figure}
\centering
\includegraphics[width=1.0\linewidth]{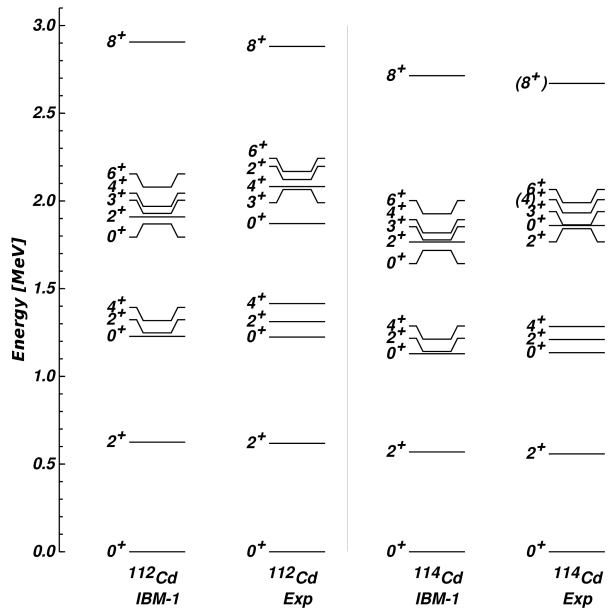}
\caption{Theoretical and experimental partial level schemes of $^{112,114}$Cd. 
The experimental data are taken from \cite{nndc}.}
\label{110_112Pd_schemes}
\end{figure}

The eigen states and eigen values were calculated with the program package PHINT
\cite{phint}. The model parameters, obtained from a fit to experimental data, 
are summarized in Table~\ref{IBM-1_par}. Experimental level energies and $B(E2)$ 
values were used for the purpose of the fit. $^{112,114}$Cd's 
theoretical and experimental level schemes are compared in 
Fig.~\ref{110_112Pd_schemes}.

\begin{table}
\begin{centering}
\caption[]{IBM-1 parameters for $^{112,114}$Cd.
\label{IBM-1_par}}
\begin{tabularx}{0.5\textwidth}{@{}l *5{>{\centering\arraybackslash}X}@{}}
\hline
\hline
\\[-1em]
Isotope & $\varepsilon$ & $\kappa$ & $\kappa'$ & $\chi$ \\
\hline
\\[-1em]
$^{112}$Cd& 0.66 & 0.0065 & -0.005  & -0.089 \\
$^{114}$Cd& 0.63 & 0.0075 & -0.005  & -0.089 \\

\hline
\hline
\end{tabularx}
\end{centering}
\end{table}

The explicit form of the $E2$ transition operator is
\begin{equation}
\label{BE2}
T(E2)=e_B[(s^\dagger\widetilde{d}+d^\dagger s)
+\chi (d^{\dagger}\widetilde{d})^{(2)}]=e_BQ ,
\end{equation}
where $e_B$ is the effective bosonic charge. The reduced transition probabilities 
are then calculated as 

\begin{equation}
\label{BsL}
B(E2; J_i \rightarrow J_f)=\frac{1}{2J_i +1} 
\left.  \left. \langle \left. J_f \| T(E2) \right. \| 
J_i \right. \rangle  \right. ^{2}.
\end{equation}
Here, $J_i$ and $J_f$ denote the spins of the initial and final state, 
respectively. The effective bosonic charge $e_B$~=~0.103~eb is adopted. It is 
determined from the experimental $B(E2)$ value for the  $2_1^+ \rightarrow 0_1^+$ 
transition in $^{112}$Cd. Theoretical and experimental transition probabilities 
for several low-lying transitions in $^{112,114}$Cd are presented in 
Table~\ref{IBM_trans_prob}. A good overall agreement with the experimental data
is achieved.

\begin{table}
\begin{centering}
\caption[]{Theoretical and experimental $B(E2)$ values for transitions connecting 
several normal-parity states in $^{112,114}$Cd. The experimental data are taken 
from \cite{nndc}.\label{IBM_trans_prob}}
\begin{tabularx}{0.5\textwidth}{@{}l *7{>{\centering\arraybackslash}X}@{}}
\hline
\hline
\\[-1em]

Isotope&$E_{level}$  &$J_{i}^{\pi}$& $E_{\gamma}$ & $J_{f}^{\pi}$& B(E2)$_{exp}$           
& B(E2)$_{th}$     \\ 
       &        [keV]&             &   [keV]      &              & [W.u.]
&  [W.u.]           \\  
    
\hline
\\[-1em]                                                                       
$^{112}$Cd& 617.518  & 2$^{+}$  & 617.518  & 0$^{+}$  & 30.31 (19)  & 30.54 \\ 
$^{112}$Cd& 1224.341 & 0$^{+}$  & 606.821  & 2$^{+}$  & 51 (14)     & 46    \\      
$^{112}$Cd& 1312.390 & 2$^{+}$  & 694.872  & 2$^{+}$  & 39 (7)      & 52    \\      
          &          &          & 1312.36  & 0$^{+}$  & 0.65 (11)   & 0.001 \\          
$^{112}$Cd& 1415.480 & 4$^{+}$  & 798.04   & 2$^{+}$  & 63 (8)      & 52    \\ 

 & & & & & & \\

$^{114}$Cd& 558.456  & 2$^{+}$  & 558.456  & 0$^{+}$  & 31.1 (19)   & 35.7  \\ 
$^{114}$Cd& 1134.532 & 0$^{+}$  & 576.069  & 2$^{+}$  & 27.4 (17)   & 52    \\      
$^{114}$Cd& 1209.708 & 2$^{+}$  & 651.256  & 2$^{+}$  & 22 (6)      & 61    \\      
          &          &          & 1209.713 & 0$^{+}$  & 0.48 (6)    & 0.001 \\ 
          &          &          & 75.177   & 0$^{+}$  & 3.4 (7)     & 0.07  \\          
$^{114}$Cd& 1283.739 & 4$^{+}$  & 725.298  & 2$^{+}$  & 62 (4)      & 61    \\ 
$^{114}$Cd& 1990.3   & 6$^{+}$  & 706.6    & 4$^{+}$  & 119 (15)    & 78    \\
$^{114}$Cd& 2669.3   & 8$^{+}$  & 678.2    & 6$^{+}$  & 85 (25)     & 86    \\
\hline
\hline
\end{tabularx}
\end{centering}
\\
\end{table}

\begin{figure*}
\centering
\includegraphics[width=0.95\linewidth]{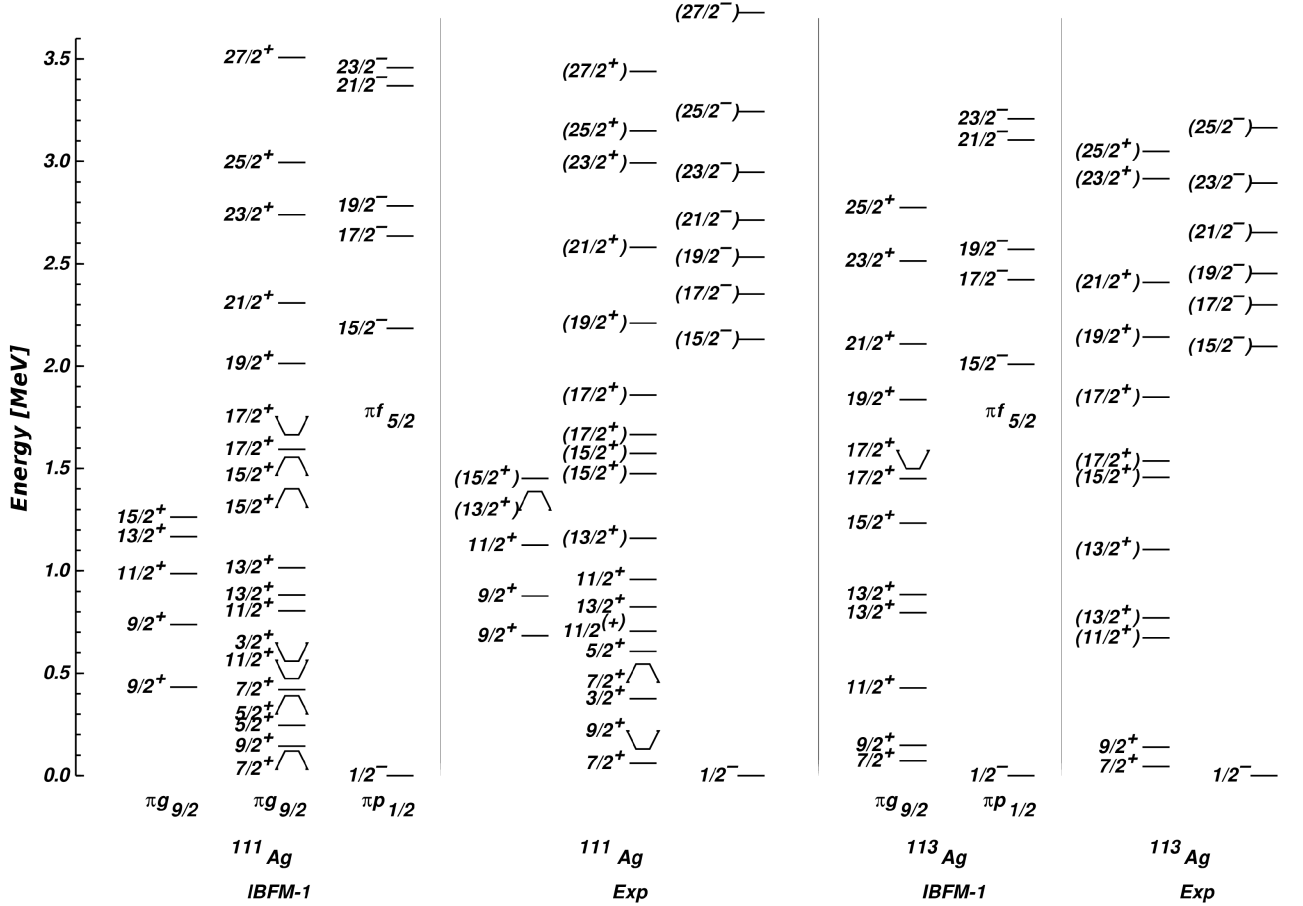}
\caption{Partial experimental and theoretical levels in $^{111,113}$Ag. Each 
sequence of states has the same leading single-particle component obtained from 
the IBFM-1 calculations. The model calculations provide information about many 
other states. A good agreement with the experimental counterparts of those 
states was observed.}
\label{111_113Ag_schemes}
\end{figure*}

\subsection{IBFM-1 calculations}
After obtaining the core eigen states, the excited states in $^{111,113}$Ag were 
calculated within the IBFM-1 model, where the Hamiltonian can be written as:
\begin{equation}
H=H_{B}+H_{F}+V_{BF},
\label{ibfmham}
\end{equation}
where H$_{B}$ is the IBM-1 bosonic Hamiltonian of the even-even core, while 
the fermionic part of the Hamiltonian  is
\begin{equation}
H_{F}=\sum_{j} E_j n_j .
\end{equation}
Here, $E_{j}$ denotes the quasiparticle energies of the single-particle shell 
model orbitals.

\begin{table}
\begin{centering}
\caption[]{BCS values for the occupation probabilities $\upsilon^2_j$
and quasiparticle energies $E_j$ of the orbitals. The same set of parameters is
used for the two nuclei. \label{BCS_table}}
\begin{tabular}{ccccccc}
\hline\hline
\hspace*{0.5cm}& $\varepsilon_j$[MeV] &$\upsilon^2_j$ & $E_j$ \\  
\hline
\\[-1em]
$\pi p_{3/2}$\hspace*{0.8cm} & 0.0  & 0.94 & 3.14 \\

$\pi f_{5/2}$\hspace*{0.8cm} & 0.4  & 0.92 & 2.80 \\

$\pi p_{1/2}$\hspace*{0.8cm} & 1.9  & 0.75 & 1.73 \\

$\pi g_{9/2}$\hspace*{0.8cm} & 1.8  & 0.77 & 1.78 \\

$\pi d_{5/2}$\hspace*{0.8cm} & 5.0  & 0.08 & 2.69 \\
\hline
\hline
\end{tabular}
\end{centering}
\end{table}
In eq.~\ref{ibfmham}, the term
\begin{equation}
\begin{multlined}
V_{BF}=\sum_{j} A_j n_d n_j + \sum_{jj'} \Gamma_{jj'} 
(Q\cdot (a_j^\dagger\widetilde{a}_{j'})^{(2)}) \\
+\sum_{jj'j''}\Lambda_{jj'}^{j''}:((d^\dagger\widetilde{a}_{j})^{(j'')}\times 
(\widetilde{d}a_{j'}^\dagger)^{(j'')})_0^{(0)}: 
\end{multlined}\label{intham}
\end{equation}
represents the boson-fermion interaction \cite{Ia79,Jo85}. This interaction
has indeed a large set of parameters which, by using 
microscopic arguments \cite{Sc80}, can be further reduced to
\begin{equation}
\begin{gathered}
A_j=A_0 , \\
\Gamma_{jj'}=\Gamma_0(u_ju_{j'}-\upsilon_j\upsilon_{j'})\left \langle j 
\left \|  Y^{(2)}\right \| j'\right \rangle , \\
\Lambda_{jj'}^{j''}=-2\sqrt{5}\Lambda_0\beta_{jj''}\beta_{j''j'}/(2j''+1)^{1/2}
(E_j + E_{j''} - \hbar\omega) .
\end{gathered}
\end{equation}
Here, 
\begin{equation}
\begin{gathered}
\beta_{jj'}=\left \langle j \left \| Y^{(2)} \right \| j' \right \rangle 
(u_j\upsilon_{j'}+\upsilon_ju_{j'}) , \\
u_j^2=1-\upsilon_j^2 ,
\end{gathered}
\label{IBFM_bf_par_b}
\end{equation}
and $\upsilon^2_j$ are the occupation probability numbers for the 
single-particle orbits $j$. Thus, $A_0$, $\Lambda_0$ and $\Gamma_0$ remain the 
only free parameters. 

The excited states in $^{111,113}$Ag were calculated by using the program package 
ODDA \cite{odda}. The single-particle energies were calculated according to the 
approach described in Ref.~\cite{Re70}. Then, they were applied to the BCS 
calculation in order to determine the respective occupation probabilities and 
quasiparticle energies, given in Table~\ref{BCS_table}. The pairing gap was set 
to $\Delta$~=~1.5~MeV. 

In the present work, the same set of boson-fermion interaction parameters 
$A_{0}$~=~-0.3~MeV, $\Gamma_{0}$~=~0.2~MeV, and $\Lambda_{0}$~=~3.8~MeV$^{2}$, 
was used to derive both the positive- and the negative-parity states in the two 
nuclei. The energy spectra, calculated for $^{111,113}$Ag are compared to their 
experimental counterparts in Fig.~\ref{111_113Ag_schemes}. The sets of levels 
having the same leading single-particle component are organized in labeled 
sequences.

In addition to the level energies, $M1$ and $E2$ transitions probabilities  
were also calculated and compared to existing experimental data. In IBFM-1,
the explicit form of the $M1$ and $E2$ operators is:
\begin{equation}
\begin{multlined}
T(M1)=\sqrt{\frac{90}{4\pi}}\mathrm{g}_d(d^{\dagger}\widetilde{d})^{(1)} \\
-\mathrm{g}_F\sum_{jj'}(u_ju_{j'}+\upsilon_j\upsilon_{j'})\cdot 
\left. \langle \left. j \| \mathrm{g}_ll+\mathrm{g}_ss \right. \| j' 
\right. \rangle \\
\times [(a_j^{\dagger}\widetilde{a}_{j'})^{(1)}+c.c.] , 
\end{multlined}
\label{IBFM_BM1}
\end{equation}

\begin{equation}
\begin{multlined}
T(E2)=e_B((s^{\dagger}\widetilde{d}+d^{\dagger}s)^{(2)}
+\chi(d^{\dagger}\widetilde{d})^{(2)}) \\
-e_F\sum_{jj'}(u_ju_{j'}-\upsilon_{j}\upsilon_{j'})
\left. \langle \left. j \| Y^{(2)} \right. \| j' \right. \rangle \\
\times [(a_j^{\dagger}\widetilde{a}_{j'})^{(2)}+c.c.] .
\end{multlined}
\label{IBFM_BE2}
\end{equation}

Here the effective bosonic and fermionic charges are denoted as $e_B$ and $e_F$ 
and were set equal to the effective bosonic charge, obtained from the respective 
even-even core. 
\begin{table*}
\begin{centering}
\caption[]{Calculated magnetic moments and transition strengths in $^{111,113}$Ag.
\label{MmTsBr}}
\begin{tabularx}{0.95\textwidth}{@{}l *7{>{\centering\arraybackslash}X}@{}}
\hline\hline
\\[-1em]
nucleus    & $\mu (7/2^+)$ & $\mu (9/2^+)$ & $B(M1)$                 & $B(E2)$                  &
$B(E2)$ & $B(E2)$ \\ 
           &               &               &$9/2^+\rightarrow 7/2^+$ & $9/2^+\rightarrow 7/2^+$ &
$11/2+->9/2+$ & $13/2+->9/2+$\\
        & $\mu _N$ & $\mu _N$ & W.u. & W.u. & W.u. & W.u. \\
\hline \\[-1em]
$^{111}$Ag &  4.81 & 5.90 & 0.028 & 27 & 43  &  6\\
$^{113}$Ag &  4.71 & 5.80 & 0.018 & 41 & 59  &  5\\
\hline\hline
\end{tabularx}
\end{centering}
\end{table*}

Given that no experimental data for $M1$ or $E2$ transitions are available in 
$^{113}$Ag, only $^{111}$Ag will be considered. In $^{111}$Ag, the half-life of 
the $9/2^+$ state is 1.22~(2)~ns and the $9/2^+\rightarrow 7/2^+$ transition is 
known to be of a mixed $M1+E2$ multipolarity with a mixing ratio of 
$\delta \leq 0.12$. The experimental $B(M1;9/2^+\rightarrow 7/2^+)=0.024$ W.u. 
and hence is hindered with respect to the single particle estimates. Since the
$E2$ transition strength strongly depends on $\delta$, estimations based on the 
mixing ratios in the lighter $^{105,107}$Ag isotopes were made. They lead to 
$B(E2)=120$ (70) W.u. for this particular transition, suggesting that collective 
modes are involved 
to a large extent. On the other hand, the hindrance of $B(M1)$ suggests that
either $\Delta L=2$ single-particle orbits are involved, or a more complex
configuration exists in the silver spectra at low energies. The only positive 
parity orbit placed close to the Fermi surface is the intruder $\pi g_{9/2}$ 
level and, hence, the scenario for $l$-forbidden transition can be ruled out. 
Indeed, the structure of the two states, which appear in all neutron mid-shell 
Ag nuclei, is considered to arise from a $\pi g_{9/2}^{-3}$ configuration.
In the seniority scheme framework, however, the $M1$ transition would be 
forbidden \cite{PVI}, which vaguely agrees with the experimental observable.
Even though such three-particle configurations are outside the IBFM-1 model
space, it is worth pushing the model to the limit and test it for the two Ag 
isotopes. 

In the present work, the $d$-boson $\mathrm{g}$-factor 
$\mathrm{g}_d$~=~0.3~$\mu_N$ was determined from the magnetic moment of the 
first 2$^+$ state in the neighbouring Cd nuclei \cite{nndc}, where 
$\mathrm{g}_s$~=~4.0~$\mu_N$ and $\mathrm{g}_l$~=~1.0 were used. By using this 
value, and the IBFM-1 Hamiltonian parameters deduced from the level energies, we 
obtain $B(M1; 9/2_{1}^{+} \rightarrow 7/2_{1}^{+})$~=~0.028~W.u. which is 
consistent with the experimental value of 0.024~(1)~W.u.. The theoretical 
calculations show that the $9/2_1^+ \rightarrow 7/2_1^+$ transition has a 
collective component, with $B(E2)$~=~27~W.u., which vaguely agrees with the 
strength of 120 (70) W.u. estimated from the experimental data. This discrepancy 
suggests that precise experimental measurement of $M1+E2$ mixing ratio is needed 
before making any firm conclusion on the nature of the states involved. Magnetic 
moments for $7/2^+$ and $9/2^+$ states are calculated and presented in 
Table~\ref{MmTsBr} along with transition strengths for the 
$9/2_1^+ \rightarrow 7/2_1^+$, $11/2_1^+ \rightarrow 9/2_1^+$ and 
$13/2_1^+ \rightarrow 9/2_1^+$ transitions. Given that there are no lifetime 
measurements available for the higher-lying excited states in 
$^{111,113}$Ag only branching ratios can be extracted from the data and used as 
a reference to the theoretical calculations. They are presented in Table~\ref{BR}, 
where a good agreement with the experimental data is observed.

\begin{table}
\begin{centering}
\caption[]{Experimental and Theoretical Branching Ratios in $^{111}$Ag and $^{113}$Ag.
\label{BR}}
\begin{tabular}{lllllll}
\hline\hline
nucleus     &              &  IBFM & Exp \\ 
\hline \\
 $^{111}$Ag &  $11/2^+\rightarrow 9/2^+$ & 100   & 100.0 (14) \\
 $^{111}$Ag &  $11/2^+\rightarrow 7/2^+$ &  0.1  &  6.0 (8) \\
 $^{111}$Ag &  $13/2^+\rightarrow 9/2^+$ &  100  &  100.0 (17)\\
 $^{111}$Ag &  $13/2^+\rightarrow 11/2^+$&  15.3 &  5.0 (17)\\
$^{113}$Ag  &  $11/2^+\rightarrow 9/2^+$ &  100  &  100  \\
$^{113}$Ag  &  $11/2^+\rightarrow 7/2^+$ &  0    &  0    \\
$^{113}$Ag  &  $13/2^+\rightarrow 9/2^+$ &  100  &  100.0 \\
$^{113}$Ag  &  $13/2^+\rightarrow 11/2^+$&  12.8 &  28.5 \\
\hline\hline
\end{tabular}
\end{centering}
\end{table}

\begin{table*}
\begin{centering}
\caption[]{Contributions of different single-particle components to the 
wave functions of the lowest-lying IBFM-1 states in $^{111,113}$Ag nuclei.
\label{111_113Ag_wf}}
\begin{tabularx}{0.75\textwidth}{@{}l *7{>{\centering\arraybackslash}X}@{}}
\hline\hline
\\[-1em]
 & & & $^{111}$Ag & &  & \\ 
\hline \\[-1em]
$J^{\pi}$& $E_{level}^{\dagger}$ [keV] & $\pi p_{3/2}$ [\%] & $\pi f_{5/2}$ [\%]& $\pi p_{1/2}$ [\%]& $\pi g_{9/2}$ [\%] & $\pi d_{5/2}$ [\%] \\
\hline
\\[-1em]
1/2$^{-}$& 0.0  & 2.9778  & 6.3163     & 90.7059   & 0.0   & 0.0  \\
7/2$^{+}$& 60   & 0.0  &  0.0  & 0.0   & 97.4192   & 2.5808  \\
9/2$^{+}$& 130  & 0.0  &  0.0  & 0.0   & 97.9084   & 2.0916  \\
\hline
 & & & $^{113}$Ag & &  & \\ 
\hline \\[-1em]
$J^{\pi}$& $E_{level}^{\dagger}$ [keV] & $\pi p_{3/2}$ [\%] & $\pi f_{5/2}$ [\%]& $\pi p_{1/2}$ [\%]& $\pi g_{9/2}$ [\%] & $\pi d_{5/2}$ [\%] \\
\hline
\\[-1em]
1/2$^{-}$& 0.0    & 3.6437  & 7.7439   & 88.6123  & 0.0 & 0.0  \\
7/2$^{+}$& 44     & 0.0  & 0.0  & 0.0  & 97.2490  & 2.7510  \\
9/2$^{+}$& 139    & 0.0  & 0.0  & 0.0  & 97.1381  & 2.8619  \\
\hline
\hline
$^{\dagger}$from \cite{nndc}
\end{tabularx}
\end{centering}
\end{table*}

Although the structure of the low-lying states in Ag nuclei is rather complex 
and involves degrees of freedom that are outside of the present theoretical 
approach, some conclusions can be drawn from the present work. The core nuclei  
$^{112,114}$Cd are well reproduced by the IBM-1 calculations. Their level schemes 
are typical for the U(5) nuclei \cite{Bo88}, where the relative position 
of the $0_2^+$ with respect to the $2^+,4^+$ doublet from the second phonon is an 
indicative feature.

A summary of the strongest single-particle contributions to the wave functions of 
the lowest-lying states in $^{111}$Ag and $^{113}$Ag is presented in 
Table~\ref{111_113Ag_wf}. The analysis shows that $\pi g_{9/2}$ has a major 
contribution to the lowest-lying positive-parity states. The leading configuration 
for the ground states in $^{111}$Ag and $^{113}$Ag is $\pi p_{1/2}$. This 
configuration is responsible also for other low-lying negative-parity states, 
not shown in Fig.~\ref{111_113Ag_schemes}. At higher energies, however, the $\pi f_{5/2}$ 
starts to play a role in the structure of the negative-parity states. In particular, 
this is the case for all negative-parity states with energies above $\sim$2 MeV,
shown in Fig.~\ref{111_113Ag_schemes}. All negative parity states with 
$J^\pi \ge 15/2^-$ have their counterparts in the experimental data. It is 
interesting to note that the $15/2^-$ band-head energy is well reproduced, but 
the experimental sequence on top of it shows a more rotational behavior, 
suggesting that the experimental bands have a quadrupole deformation higher than 
that of the respective Ag ground state.

\section{Conclusions}
$^{111,113}$Ag nuclei were populated in induced fission reactions. The level 
schemes were extended and the spin/parities of the new levels are based on
systematics. To interpret the structure of the excited states IBFM-1 was used.
The ground state in the two nuclei is associated to the $\pi p_{1/2}$ 
configuration. At higher energies, above 2 MeV, the $\pi f_{5/2}$ configuration
starts to play a leading role. Those levels are arranged in band-like sequence,
suggesting a deformation, larger than the ground state deformation, is developed 
there. The positive-parity states are also well described by the model. The 
$(7/2^+,9/2^+)$ doublet splitting is correctly reproduced. The 
$B(M1; 9/2^+\rightarrow 7/2^+)$ transition rate in $^{111}$Ag agree with the 
experimental observations, while further experimental data on the $M1+E2$ 
mixing ratio is needed in order to have a better experimental reference point. 
However, an overall good agreement with the experimental data is observed, 
suggesting that the $^{111,113}$Ag nuclei exhibit well-developed collective 
properties.

\section{Acknowledgements}
This work is supported by the Bulgarian National Science Fund under contract
number DFNI-E02/6 and by the German BMBF under grant number 06BN109. SK 
appreciates the number of interesting and constructive discussions with 
D.Bucurescu regarding the IBFM-1 calculations. SL acknowledges the constructive 
discussions with P. Van Isacker and V. Paar.

\section{Addendum}
While preparing the present manuscript, $^{113}$Ag partial level scheme was 
observed from an experiment performed at GANIL and published in Ref.\cite{Ki17}

\end{document}